\documentclass[
reprint,
superscriptaddress,
amsmath,amssymb,
aip,
apl,
floatfix
]{revtex4-2}

\usepackage{graphicx}
\usepackage{dcolumn}
\usepackage{bm}
\usepackage[colorlinks,citecolor=magenta]{hyperref}

\begin{document}

\title{Twisted heterobilayer photonic crystal based on stacking and selective etching of 2D materials}

\author{Qing Wang}
\affiliation{Institute of Physics, Chinese Academy of Sciences, Beijing 100190, China}
\affiliation{School of Physics, Liaoning University, Shenyang 110036, PR China}
\author{Yuhang Li}
\affiliation{Institute of Physics, Chinese Academy of Sciences, Beijing 100190, China}
\affiliation{School of Physical Sciences, University of Chinese Academy of Sciences, Beijing 100049, China}
\author{Shaofeng Wang}
\affiliation{Institute of Physics, Chinese Academy of Sciences, Beijing 100190, China}
\affiliation{School of Physics, Liaoning University, Shenyang 110036, PR China}
\author{Shuo Cao}
\email{shuocao@lnu.edu.cn}
\affiliation{School of Physics, Liaoning University, Shenyang 110036, PR China}
\author{Xiulai Xu}
\email{xlxu@pku.edu.cn}
\affiliation{State Key Laboratory for Mesoscopic Physics and Frontiers Science Center for Nano-optoelectronics, School of Physics, Peking University, 100871 Beijing, China}
\affiliation{Peking University Yangtze Delta Institute of Optoelectronics, Nantong, Jiangsu 226010, China}
\affiliation{Collaborative Innovation Center of Extreme Optics, Shanxi University, Taiyuan, Shanxi 030006, China}
\author{Chenjiang Qian}
\email{chenjiang.qian@iphy.ac.cn}
\affiliation{Institute of Physics, Chinese Academy of Sciences, Beijing 100190, China}
\affiliation{School of Physics, Liaoning University, Shenyang 110036, PR China}

\begin{abstract}
Nanophotonic devices with moiré superlattice is currently attracting broad interest due to the unique periodicity and high efficiency control of photons.
Till now, experimental investigations mainly focus on the single layer device, i.e., two or more layers of photonic crystal patterns are merged and etched in a single layer of material.
By comparison, twisted photonic crystal with multilayer materials raises challenges in the nanofabrication technology, because the growth of upper layer material usually requires a smooth bottom layer without nanostructures.
Hereby, we fabricate twisted heterobilayer photonic crystal in the graphite/Si$_3$N$_4$ heterostructure.
We use dry transfer method to stack the graphite on top of bottom Si$_3$N$_4$ with pre-etched photonic crystal patterns.
Selective dry etching recipes are used to etch two photonic crystal layers individually, which improves the quality and accuracy in alignment.
The cavity photonic mode at the visible wavelength $\sim 700$ nm arsing from the moiré site is clearly observed in experiment.
These results reveal the experimental diagram of heterobilayer nanophotonic devices and open the way to design flexibility and control of photons in new degrees of freedom.
\end{abstract}

\maketitle

\section{Introduction}

Twisted photonic crystal (PHC) exhibit unique interference effects arising from their periodic moiré superlattice \cite {10.1038/s41563-024-01950-9, 10.1063/5.0070163}.
With optimized twist angles, localized flat bands appear within the photonic bandgap and significantly slow down the light propagation, introducing cavity photonic modes periodic in spatial \cite{10.1038/s41586-019-1851-6, 10.1021/acsphotonics.3c01064}.
These modes have been optimized for an optimal combination of high quality factor (Q-factor) and small mode volumes compared to conventional PHC cavities \cite{10.1063/5.0105365, 10.1038/s41565-021-00956-7}.
Precise control over the mode profile wavelength and spatial distribution by tuning the twist angle provides new opportunities for on-chip photonic devices and tunable optical states.
These advantages indicate twisted PHC as promising platform for applications in low threshold laser \cite{10.1038/s41565-021-00956-7, 10.1126/sciadv.ade8817, 10.1038/s41586-023-06789-9}, nonlinear optics \cite{10.1002/lpor.202000596, 10.1021/acs.nanolett.4c03632}, and high efficiency quantum sources \cite{10.48550/arxiv.2411.16830}.

Till now, most experiments on twisted PHC are implemented with single layer devices  \cite{10.1038/s41563-024-01950-9, 10.1088/1361-6528/ad024a}.
This means the two or more layers of PHC patterns for the moiré superlattice are merged and etched in a single layer of material as a 2D system.
This approach offers a good compatibility with current fabrication technologies for 2D slab photonic crystals which have been optimized for decades \cite{10.1038/nature02063,10.1117/1.OE.62.1.010901}.
However, such single layer structure limits the design flexibility of devices and controllability of photons, e.g., in this case the flat photonic band only exists with certain "magic" twist angles \cite{10.1038/s41565-021-00956-7, 10.1002/lpor.202000596}.
In contrast, twisted multilayer PHC as a 3D system introduce the design flexibility in the material, the thickness, and the gap between different layers \cite{10.1038/s41377-021-00601-x, 10.1103/PhysRevLett.126.223601}.
These features have been predicted to allow the high controllability of photons through the flat band with different twist angles \cite{10.1103/PhysRevLett.126.223601}, and moreover, enable the control of photons propagating along the vertical axis such as the generation of optical vortex \cite{10.1038/s41467-023-41068-1, 10.1126/science.1223824, 10.1002/adom.201600629}.

\begin{figure*}[htbp]
\centering
\includegraphics[width=\linewidth]{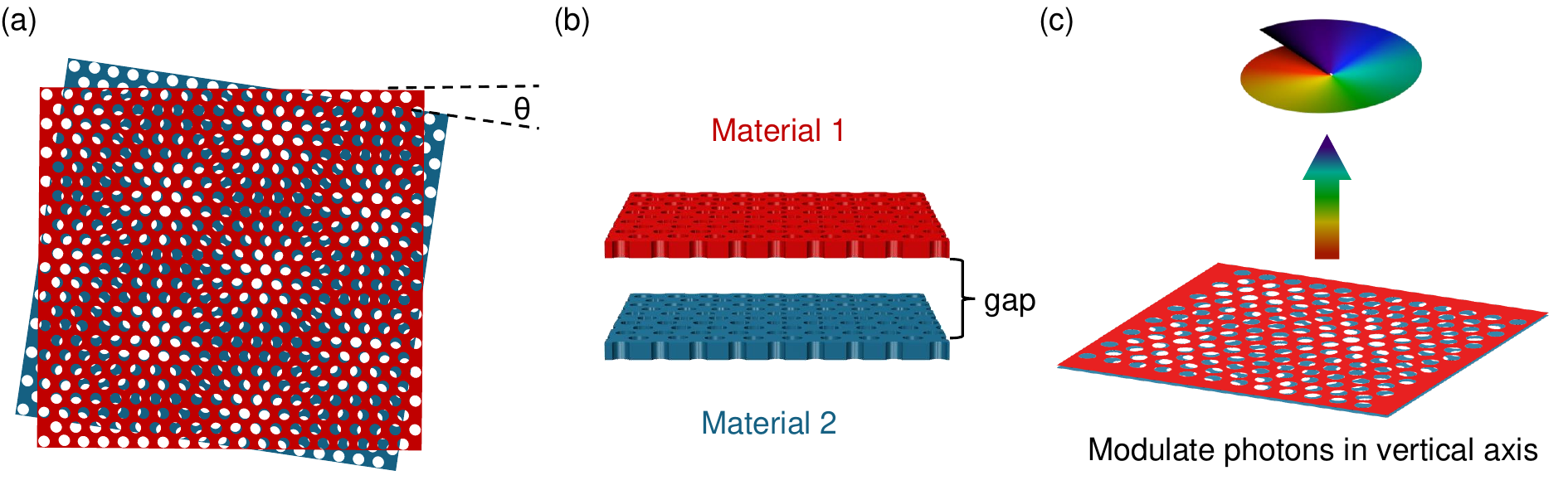}
\caption{
    Schematic of twisted heterobilayer PHC.
    (a) Top (red) and bottom (blue) layers of PHC slabs with a twist angle $\theta$ exhibit the periodic moiré superlattice.
    (b) Compared to the single layer device, the heterobilayer introduces new degrees of freedom in the materials and the gap between different layers.
    (c) The heterobilayer as a 3D system enables the control of photons propagating along the vertical axis.
}
\label{f1}
\end{figure*}

The challenge of twisted multilayer photonic structure mainly lies in fabrication technologies.
One approach is assembling two layers of fabricated PHC slabs.
For example, H. Tang et al. reported the homobilayer PHC by assembling two Si$_3$N$_4$ slabs and demonstrated the moiré scattering in the NIR range \cite{10.1126/sciadv.adh8498}.
While the assemble method works well for long wavelength devices having the $\mathrm{\mu}$m-scale spatial extent, e.g., PHC lattice constant and slab thickness, devices at visible wavelength regime usually have the spatial extent around 100 nm scale and thus have much less mechanical stability.
Another approach is alternately etching the PHC patterns in one layer and then growing the upper layer material.
The challenge in this approach is that the high quality growth of the conventional hard membrane material such as Si or GaAs generally requires a smooth bottom layer without nanostructures \cite{10.1021/acs.nanolett.5b04448,10.1364/ome.6.002360}.
Nonetheless, 2D materials including graphite, hexagonal boron nitride, and transition metal dichalcogenides are soft and can be stacked as heterostructures using the dry transfer method \cite{10.1126/science.aac9439,10.1002/adma.201804828}.
Recent works have reported that 2D materials can be transferred on top of substrate with nanostructures and serve as a functional dielectric component, resulting in ultra high Q nanophotonic cavities embedded with 2D heterostructures \cite{10.1103/PhysRevLett.128.237403, 10.1126/sciadv.adk6359}.
These unique properties indicate the potential in 3D photonic system by combing the conventional 2D nanofabrication and the stacking in out-of-plane direction.

In this work, we fabricate twisted heterobilayer PHC in the graphite/Si$_3$N$_4$ heterostructure.
The first layer PHC pattern is etched in the Si$_3$N$_4$ using e-beam lithography (EBL) and inductively coupled plasma etching (ICPRIE).
The graphite flake is then transferred on top of the Si$_3$N$_4$, following by the selective etching of second layer PHC pattern in which the recipe does not affect the first Si$_3$N$_4$ layer.
By controlling the twist angle, corresponding moiré superlattices are successfully achieved.
The accurate alignment is demonstrate in scanning electron microscope (SEM) images, and the moiré cavity photonic mode at the visible wavelength $\sim 700$ nm is further demonstrated by the position-dependent photoluminescence (PL) spectroscopy.
The heterobilayer structure not only expands the design flexibility of moiré photonic devices, but also opens the way to interlayer photonic coupling and control of photons in new degrees of freedom.

\section{Results}

The schematic of the twisted heterobilayer PHC is presented in Fig. \ref{f1}(a), including the first bottom layer denoted by the blue color and the second top layer denoted by the red color.
When the twist angle $\theta$ exists between the top and bottom layer, periodic moiré superlattice arises, exhibiting the periodic bright (AA stack) and dark (AB/BA stack) regions.
The moiré superlattice provides the confinement of photons at the energy of flat band within the AA region.
Such cavity photonic mode has been widely obtained in the single layer devices, which means the two layers of patterns are merged into one and etched in a single layer material \cite{10.1038/s41565-021-00956-7, 10.1126/sciadv.ade8817, 10.1038/s41586-023-06789-9, 10.1002/lpor.202000596, 10.1021/acs.nanolett.4c03632}.
By comparison, the two layers of patterns could be etched in two different layers, as shown in Fig. \ref{f1}(b).
The material and thickness of the two PHC layers could be different, along with the controllable gap or dielectric between them.
As such, the twisted heterobilayer PHC introduces the design flexibility in these new degrees of freedom and advances the controllability of photons.
For example, in the single layer device, the flat band could only be achieved with the "magic" twist angles which means each moiré site is perfectly uniform \cite{10.1038/s41565-021-00956-7, 10.1002/lpor.202000596}.
In contrast for the heterobilayer device, even when the twist angle is not "magic", the flat band has been predicted by controlling the material and the gap \cite{10.1103/PhysRevLett.126.223601}.
Moreover, as schematically shown in Fig. \ref{f1}(c), the heterobilayer structure paves the way to control the photons in the out-of-plane direction, such as the generation of optical vortex predicted by T. Zhang et al. \cite{10.1038/s41467-023-41068-1}.
These predicted features indicate great potential for the fabrication and experimental exploration of twisted heterobilayer PHC.

\begin{figure*}[htbp]
\centering
\includegraphics[width=\linewidth]{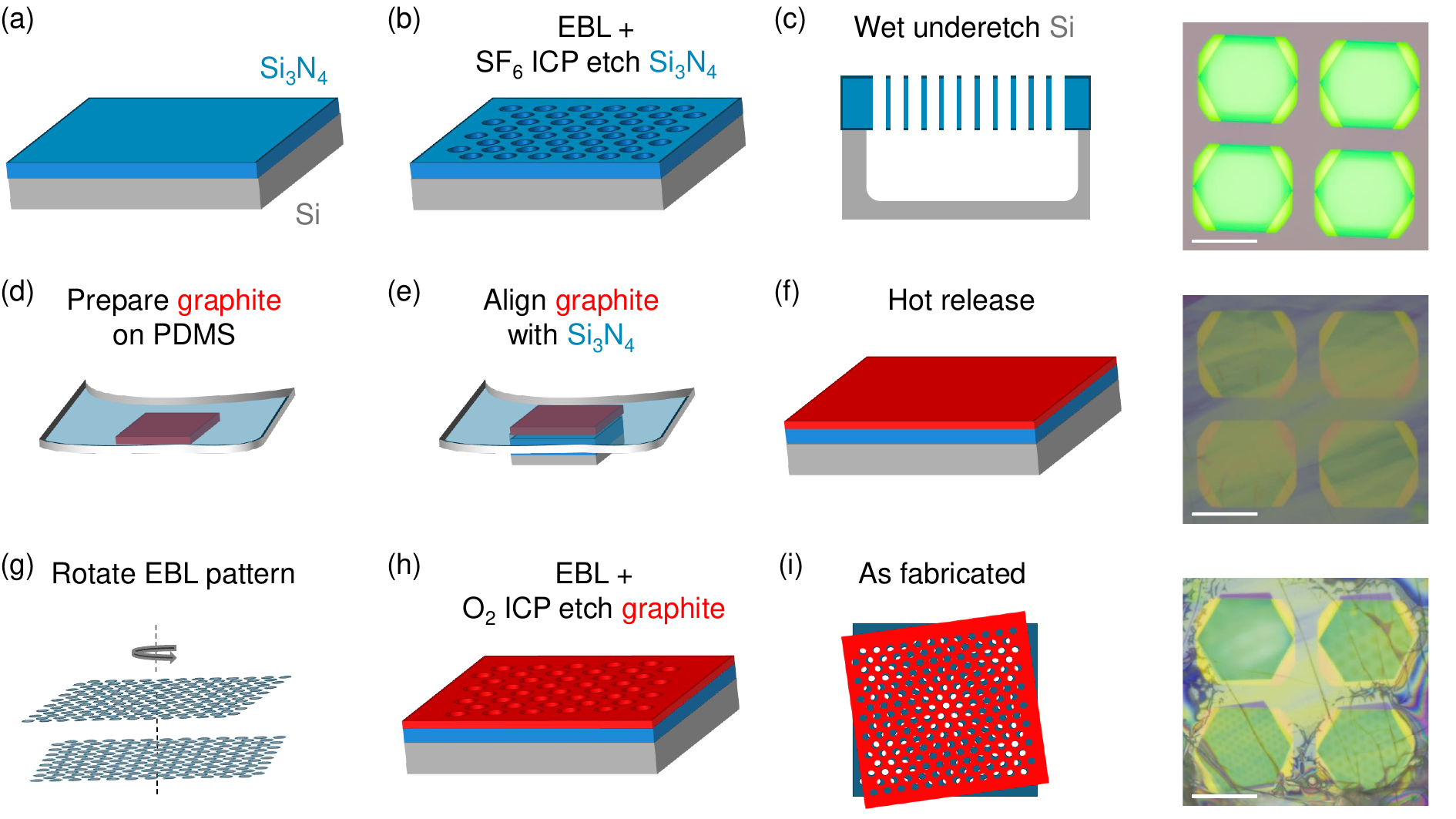}
\caption{
    Fabrication procedures.
    (a) Bare substrate with a 200 nm thick Si$_3$N$_4$ on Si.
    (b) EBL and ICPRIE etching with SF$_6$ gas are used to etch the first layer PHC pattern in Si$_3$N$_4$.
    (c) Wet underetching is used to remove the bottom Si.
    (d) Exfoliating graphite flake on the PDMS.
    (e) Aligning and assemble the graphite/Si$_3$N$_4$ heterostrcutre using dry transfer method.
    (f) Graphite flake is released on the Si$_3$N$_4$ after heating.
    (g) Rotating the PHC pattern for the second layer.
    (h) EBL and ICPRIE etching with O$_2$ gas are used to etch the PHC pattern in graphite.
    (i) Schematic of the device after fabrication.
    Right panels in (c)(f)(i) are optical microscope images after these procedures.
    The white scale bar is 20 $\mathrm{\mu}$m.
}
\label{f2}
\end{figure*}

We develop the fabrication method of twisted heterobilayer PHC based on the stacking of 2D materials.
The key procedures are presented in Fig. \ref{f2}.
The fabrication begins with the etching of first layer PHC in Si$_3$N$_4$, as shown in Fig. \ref{f2}(a)-(c).
The Si$_3$N$_4$ has the thickness of 200 nm, supporting on the Si substrate wafer with the thickness of 525 $\mathrm{\mu}$m.
The PHC pattern has the lattice constant $a$ of 330 nm and the nanoholes with the radius $r$ of 96 nm in a hexagonal lattice.
The pattern is written through EBL with an acceleration voltage of 100 keV and a dose of 360 $\mathrm{\mu C \cdot cm^{-2}}$ into the AR-P 6200 resist with a thickness of 500 nm.
The resist is used as the etching mask in the following ICPRIE etching, utilizing the SF$_6$:C$_4$F$_8$=3:2 gas mixture with the pressure of 18 mTorr and DC bias voltage of 109 V.
The KOH solution (45 wt. $\%$) is then used in the wet etching to remove the Si at bottom and create the suspended PHC structure.
After the fabrication of first layer PHC, we prepare and transfer the graphite flake on top as shown in Fig. \ref{f2}(d)-(f).
The high-quality graphite flake is prepared on PDMS using the mechanical exfoliation and has the thickness of 50 nm.
Then the dry transfer method is used to roughly align and stack the graphite on top of the first layer PHC.
Finally, we etch the second layer PHC with the twist angle into the graphite, as shown in Fig. \ref{f2}(g)-(i).
The parameters for the EBL is same as those in Fig. \ref{f2}(b), whilst a selective ICPRIE etching recipe is used in Fig. \ref{f2}(h), utilizing O$_2$ gas with the pressure of 10 mTorr and DC bias voltage of 357 V.
This recipe only etches the graphite while does not affect the bottom Si$_3$N$_4$, as discussed in detail next.
After the etching of second layer PHC, the twisted heterobilayer PHC is sucessfully fabricated as shown in Fig. \ref{f2}(i).

We emphasize that in our method, we transfer the 2D material flake (Fig. \ref{f2}(e)(f)) and then etch the second layer PHC (Fig. \ref{f2}(g)(h)), rather not in the reversed order of these two steps.
This raises challenges in the nanofabrication, i.e., in the etching of the second layer PHC we need to avoid damaging the bottom Si$_3$N$_4$ layer.
However, if we firstly etch the graphite and then transfer, the two layers would be aligned using the optical microscopy during the dry transfer for which the accuracy is around 1 $\mathrm{\mu}$m \cite{10.1021/acs.nanolett.5b05263}.
For a 20-$\mathrm{\mu}$m structure this corresponds to the accuracy in twist angle $\sim 2.9^\circ$.
Moreover, the dry transfer is expected to introduce strain in the 2D flake, and thereby, the PHC pattern would be deformed \cite{10.1088/2053-1583/ab7629, 10.1021/acsnano.4c00590}.
In contrast, our method improves the accuracy in the alignment between two PHC layers, since they are aligned by the e-beam microscopy during the EBL for which the accuracy is around 10 nm.
This corresponds to the accuracy in twist angle $\sim 0.03^\circ$.
The PHC pattern in the 2D material flake will not be deformed by the strain since it is etched after the dry transfer \cite{10.1103/PhysRevLett.128.237403}.
Therefore, our method exhibits better performance in improving the quality and alignment accuracy of twisted heterobilayer PHC.

\begin{figure}[htbp]
\centering
\includegraphics[width=\linewidth]{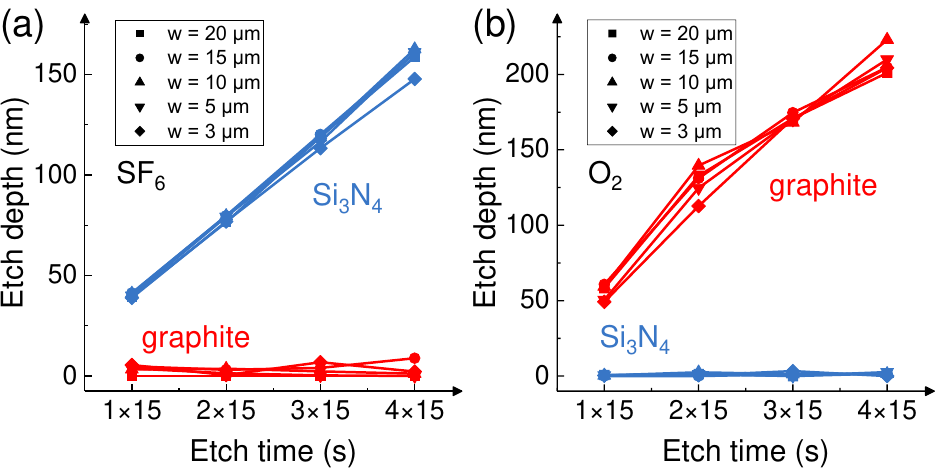}
\caption{
    Selectivity in ICPRIE etching.
    (a) Recipe with SF$_6$ gas etches Si$_3$N$_4$ whilst not graphite.
    (b) Recipe with O$_2$ gas etches graphite whilst not SI$_3$N$_4$.
    Etching rates are measured from the trenches with different width $w$.
}
\label{f3}
\end{figure}

We develop the selective dry etching recipe as discussed in the context of Fig. \ref{f2} to avoid damaging the other layer during the etching.
The selectivity of etching recipes are presented in Fig. \ref{f3}.
We etch trenches with different width $w$ using different time periods in the ICPRIE.
In Fig. \ref{f3}(a) we present the results of etching with SF$_6$ gas.
As shown, the etching depth of Si$_3$N$_4$ linearly increases with the etching time, exhibiting the etching rate $2.66 \pm 0.02\ \mathrm{nm \cdot s^{-1}}$.
In contrast, the etching of graphite with SF$_6$ gas is negligible \cite{10.1021/acsami.1c09923}.
In Fig. \ref{f3}(b) we present the results of etching with O$_2$ gas, for which the recipe has been widely applied to etch graphite \cite{10.1016/j.microrel.2012.07.013}.
In this case, the graphite is significantly etched with the rate around $3.34 \pm 0.16\ \mathrm{nm \cdot s^{-1}}$, while the Si$_3$N$_4$ is not etched.
The slight nonlinearity in the etching of graphite is attributed to a striking process at the beginning of etching.
Nonetheless, the selectivity in these two recipes are clearly demonstrated by these experimental data.

\begin{figure*}[htbp]
\centering
\includegraphics[width=\linewidth]{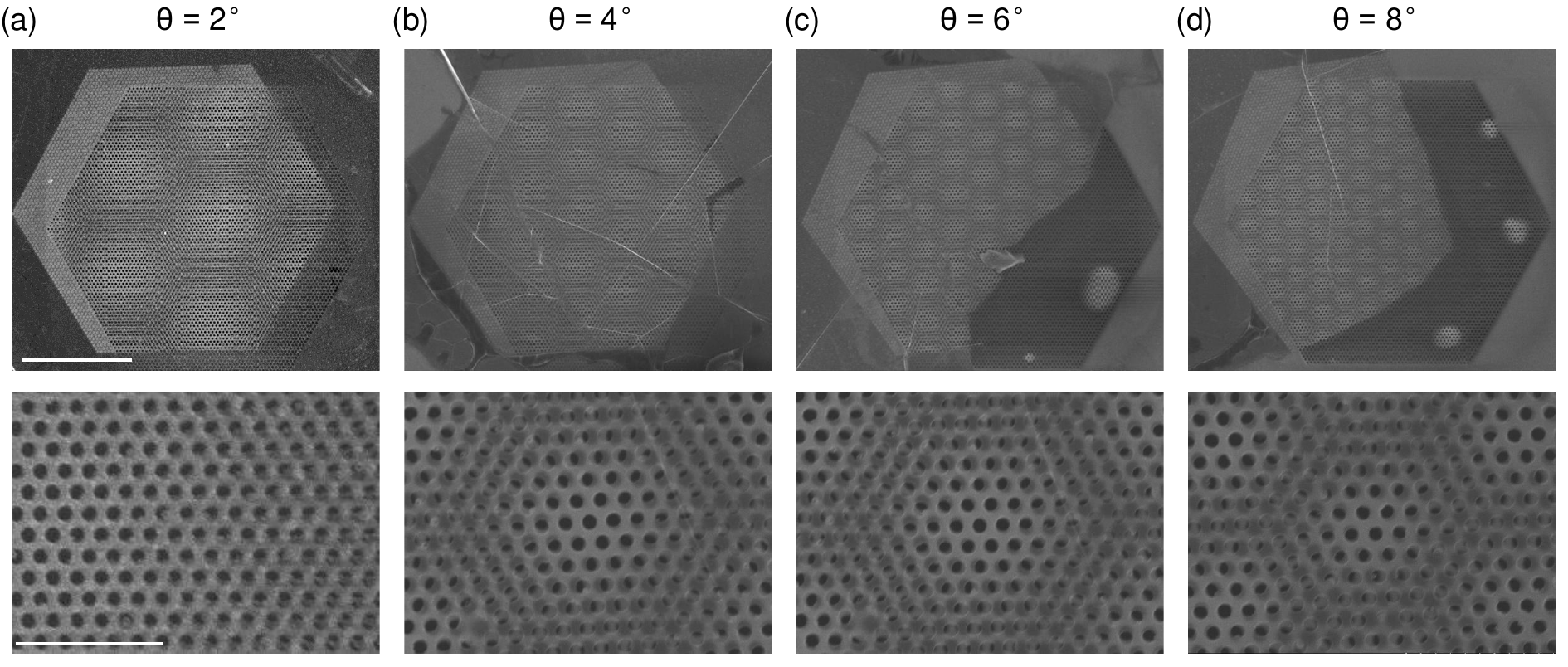}
\caption{
    SEM images of twisted heterobilayer PHC.
    The twist angle $\theta$ is (a) 2$^\circ$ (b) 4$^\circ$ (c) 6$^\circ$ (d) 8$^\circ$, correspondingly.
    Moiré superlattice with the periodicity decreasing as $\theta$ increases is clearly observed.
    The white scale bar in the upper panel is 10 $\mathrm{\mu}$m, and that in the bottom panel is 2 $\mathrm{\mu}$m.
}
\label{f4}
\end{figure*}

In experiment, we fabricate the devices with four different twist angles 2$^\circ$, 4$^\circ$, 6$^\circ$, and 8$^\circ$, and their SEM images after fabrication are presented in Fig. \ref{f4}(a)-(d), respectively.
As shown, the periodic moiré sites clearly appears, and no deformation or distortion in the upper graphite flake is observed.
In theory, the periodicity of moiré sites is $\Delta_m=a/\left[2\sin{\left( \theta/2 \right)}\right]$ \cite{10.1038/nature26154}, and the value is 9454, 4728, 3153, and 2365 nm for the four twist angles.
We extract the periodicity from the SEM images in Fig. \ref{f4} as $9450 \pm 39$, $4701 \pm 20$, $3164 \pm 14$, and $2393 \pm 11$ nm.
The values correspond to the experimental twist angle of $2.00 \pm 0.01^\circ$, $4.02 \pm 0.02^\circ$, $5.98 \pm 0.03^\circ$, $7.91 \pm 0.04^\circ$, respectively, agreeing perfectly with the designed values.
These results extracted from the SEM images demonstrate the excellent etching and accurate alignment of our method to fabricate twist heterobilayer PHC.

\begin{figure}[htbp]
\centering
\includegraphics[width=\linewidth]{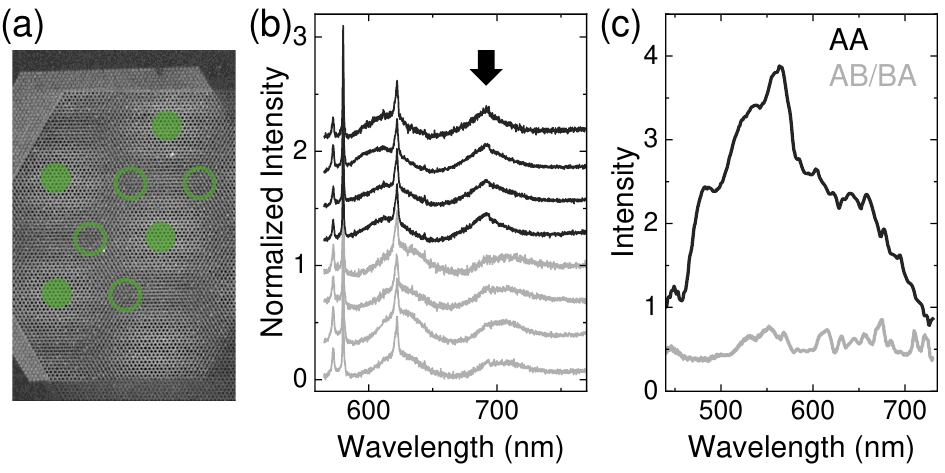}
\caption{
    PL spectroscopy of moiré sites.
    (a) We record data from four AA positions (solid green dots) and four AB/BA positions (hollow green dots) from the device with $\theta=2^\circ$.
    (b) Measured spectra in which black lines corresponding to AA and gray lines corresponding to AB/BA.
    The arrow denotes the confined cavity mode.
    (c) Theoretical calculation using FDTD method.
}
\label{f5}
\end{figure}

Finally, we implement PL spectroscopy using a micro-PL setup to identify the confinement of photons in the moiré site.
The sample is excited by a 532-nm cw-laser having a spot size $\sim 1\ \mathrm{\mu m}$ and power $\sim 5$ mW.
Such a high excitation power excites the natural defects in the Si$_3$N$_4$ and/or graphite which function as the light source to excite the cavity photonic modes \cite{10.1038/srep28326, 10.1021/acs.nanolett.2c00739, 2210.00150}.
PL signals are collected by an objective has a magnification of 100 and an NA of 0.75 and then detected by a matrix array Si CCD detector in a spectrometer with a grating having 300 grooves per mm.
The sample is mounted on a xyz-nanopositioner for the spatially resolved spectroscopy.
As shown in Fig. \ref{f5}(a), we record data from four AA positions denoted by the solid green dots as well as four AB/BA positions denoted by the hollow green dots in the device with the twist angle of 2$^\circ$, as the large moiré periodicity allows the distinguishment between different positions in spatial.
Since in the moiré superlattice photons are confined in the AA positions, and thereby, peaks arising from the cavity photonic modes are expected to be pronounced at AA positions whilst weak at AB/BA positions \cite{10.1103/PhysRevLett.126.223601,10.1038/s41377-021-00601-x}.
The experimental spectra are presented in Fig. \ref{f5}(b), in which the black lines corresponds to AA positions, and the gray lines corresponds to AB/BA positions, respectively.
As expected, we observe a peak centered at 692 nm with a FWHM $\sim 32$ nm at AA positions, which in contrast is suppressed at AB/BA positions.
The FWHM corresponds to the Q-factor of 20.
The low Q-factor is attributed to the absorption of graphite, i.e., the refractive index of graphite has both the real and imaginary components \cite{10.1364/AO.51.003250, 10.1007/s00339-017-1249-y}.
The Q-factor could be improved in future works by the application of 2D materials with no absorption, such as hexagonal boron nitride in the visible wavelength range \cite{10.1364/OL.44.003797, 2210.00150}.
To further support the observation of cavity photonic mode, we calculated the spectra from twisted heterobilayer PHC using 3D finite-difference time-domain (FDTD) method, and typical results are plotted in Fig. \ref{f5}(c).
By comparison, the cavity photonic mode is clearly observed at the AA region.
We note that in the calculation, two peaks centered around 550 and 660 nm is observed.
In experiment, the high energy peak is resonant with Raman signals from Si$_3$N$_4$ and graphite, i.e., the three sharp peaks in Fig. \ref{f5}(b) \cite{PhysRevLett.130.126901, 10.1098/rsta.2004.1454}.
The resonance behavior makes the spectra at high energies complex, and thereby, our discussion mainly focus on the low energy peak.
The good agreement between theoretical calculation and experimental observation strengthens the cavity photonic mode in the moiré sites.

\section{Conclusion}

In conclusion, we develop the fabrication method of twisted heterobilayer PHC based on the stacking and selective etching of 2D materials.
The dry transfer method allows the stacking of second layer material on top of the substrate with pre-etched PHC nanostructures.
The selective etching after dry transfer suppresses the deformation or distortion and improves the accuracy in the alignment between two layer PHCs.
Owing to these advantages, the fabricated devices perfectly agree with the designs, and the experimental identification of cavity photonic modes is consistent to the theoretical calculations.
Since the stacking and selective etching method is valid for various kinds of 2D materials, our work paves the way to more complex devices in heterostructures with multiple layers and novel functionalities.

\begin{acknowledgments}
This work is supported by the National Natural Science Foundation of China (Grants No. 12494600, 12494601, 12494603, 12474426, 62025507 and 12374181) and the Chinese Academy of Sciences Project for Young Scientists in Basic Research (Grant No. YSBR-112).
\end{acknowledgments}

\section*{Author Declarations}

\textbf{Conflict of Interest}
Authors state no conflict of interest.

\textbf{Data Availability}
The datasets generated and analyzed during the current study are available from the corresponding authors upon reasonable request.

%

\end{document}